\documentclass[conference]{IEEEtran}

\usepackage{cite}
\usepackage{amsmath,amssymb,amsfonts}
\usepackage{graphicx}
\usepackage[caption=false,font=footnotesize]{subfig}
\usepackage{booktabs}
\usepackage{array}
\usepackage{textcomp}
\usepackage{url}
\usepackage[colorlinks = true,
linkcolor = blue,
urlcolor  = blue,
citecolor = blue,
anchorcolor = blue]{hyperref}
\graphicspath{{./}}
\newcommand{\CN}{\mathcal{CN}}
\newcommand{\E}{\mathbb{E}}
\newcommand{\argmin}{\operatorname*{arg\,min}}
\newcommand{\argmax}{\operatorname*{arg\,max}}

\begin{document}

\title{Code-Domain Grouped Index Modulation for Spectrally Efficient
Spread-Spectrum Communications}

\author{\IEEEauthorblockN{Peng~Zhang, Jian~Dang, and Zaichen~Zhang}
\IEEEauthorblockA{National Mobile Communications Research Laboratory, Southeast University, Nanjing, China\\
Purple Mountain Laboratories, Nanjing, China\\
Email: \{peng\_zhang,dangjian,zczhang\}@seu.edu.cn}}

\maketitle

\begin{abstract}
Efficient exploitation of spreading resources is important for improving the
transmission efficiency of spread-spectrum communications. Code index
modulation (CIM) conveys additional information through spreading-code
indices without requiring extra radio-frequency chains. The fixed number of
code-index branches in existing CIM schemes limits information allocation
between the code-index and modulation-symbol domains as the target rate
increases. This paper proposes code-domain grouped index modulation (CGIM),
which partitions an orthogonal spreading-code bank into multiple groups and
maps the in-phase and quadrature components of one quadrature amplitude
modulation (QAM) symbol in each group onto two independent code indices.
Group-wise despreading enables parallel detection. Group-wise
maximum-likelihood (ML) and low-complexity greedy detection (GD) are developed,
with both achieving joint ML decisions under the stated conditions. A pairwise
error probability (PEP)-based bit error rate (BER) analysis is derived for
Rayleigh and Nakagami-$m$ fading, with additive white Gaussian noise (AWGN)
as a benchmark. Numerical results show that CGIM achieves better BER performance than
traditional CIM benchmark schemes at the same transmission rate, owing to a
more balanced distribution of information bits between the code-index and
modulation-symbol domains.
\end{abstract}

\begin{IEEEkeywords}
Code index modulation, grouped index modulation, spread spectrum, pairwise error
probability, bit error rate.
\end{IEEEkeywords}

\section{Introduction}
The growing exchange of sensing, control, and infotainment data in vehicular
networks places stringent demands on transmission rate and link reliability.
Spread-spectrum transmission provides robust signaling against interference
and channel impairments, while its spreading-code dimension can also be
exploited for carrying additional information. Index modulation (IM) conveys
information through both conventional modulation symbols and the indices of
selectable transmission resources~\cite{Basar2013OFDMIM,Zhang2024RDRSM,
Zhang2026FARDIM}. Accordingly, code index modulation (CIM) embeds index bits
in spreading-code selection without extra radio-frequency chains~
\cite{Kaddoum2015CIM}.

CIM exploits the orthogonality of spreading codes for index-based signaling,
providing a promising means of expanding the information-bearing dimensions
of spread-spectrum systems without introducing additional radio-frequency
chains. Generalized CIM (GCIM) assigns the in-phase and quadrature components of a
modulation symbol to independently selected spreading codes
\cite{Kaddoum2016GCIM}. Quadruple CIM (QCIM) further employs two spreading-code indices on each of
the in-phase and quadrature branches, yielding four code indices in total to
convey additional information bits in parallel~\cite{Cai2025QCIM}. However, the number of code-index branches in these schemes is fixed by their
respective mapping structures. Supporting more index branches therefore requires
a corresponding modification of the mapper and increases the dimension of the
joint detection problem. Grouped mappings provide an alternative by partitioning
the available IM resources into several lower-dimensional subspaces
~\cite{Qian2025GCIMAFDM}. For orthogonal spreading codes, different groups can be
separated by group-wise despreading, allowing the subsequent detection to be
performed independently within each group.

Based on these observations, we propose code-domain grouped index modulation
(CGIM). Its contributions are summarized as follows.
\begin{itemize}
\item We propose CGIM, which partitions a spreading-code bank into multiple groups
and decomposes the quadrature amplitude modulation (QAM) symbol in each group into
in-phase and quadrature pulse amplitude modulation (PAM) components mapped to
separate code indices. Group-wise despreading supports parallel detection over a
configurable number of code-index branches.
\item We develop group-wise maximum-likelihood (ML) and greedy detection (GD).
We prove that group-wise ML is equivalent to joint block-wise ML and, for
equal-energy codes and square QAM, that GD is ML-equivalent. Their
real-multiplication counts are also derived.
\item We derive the bit error rate (BER) performance of CGIM by characterizing the pairwise error
probabilities (PEPs) of one representative real group branch. For Rayleigh,
Nakagami-$m$, and additive white Gaussian noise (AWGN) channels, the combining-gain
moment-generating functions and a classical Gaussian $Q$-function approximation
yield closed-form approximations to the full-pair BER union bound.
\end{itemize}

The rest of this paper is organized as follows. Section~\ref{sec:system_design}
presents the CGIM system model. Section~\ref{sec:receiver} develops the ML and
GD receivers and states their equivalence results. Sections~\ref{sec:pep} and
\ref{sec:se_complexity} present the BER, spectral-efficiency, and complexity
analyses, respectively. Section~\ref{sec:results} gives the numerical results,
and Section~\ref{sec:conclusion} concludes the paper. Appendices~A and B prove
the detector-equivalence results.

\emph{Notation:} Bold lowercase and uppercase letters denote vectors and matrices,
respectively; $\mathrm j=\sqrt{-1}$, and $\mathbf I_n$ and $\mathbf0$ are the
identity matrix and all-zero vector or matrix. The operators
$(\cdot)^{\operatorname T}$, $(\cdot)^{\operatorname H}$, $\argmin$, and
$\argmax$ denote transpose, conjugate transpose, and the minimizing and maximizing
arguments. The norms $\|\cdot\|_2$ and $\|\cdot\|_{\operatorname F}$ are Euclidean
and Frobenius norms, and $|\mathcal A|$ is the cardinality of $\mathcal A$.
The operators $\Re\{\cdot\}$, $\Im\{\cdot\}$, $\E[\cdot]$, and $\Pr(\cdot)$
denote real part, imaginary part, expectation, and probability, while $Q(\cdot)$
and $d_{\operatorname H}(\cdot,\cdot)$ are the Gaussian $Q$-function and Hamming
distance. Finally, $\CN(\boldsymbol\mu,\mathbf C)$ denotes a circularly symmetric
complex Gaussian distribution.

\section{System Model}
\label{sec:system_design}

\begin{figure}[!t]
    \centering
    \includegraphics[width=\columnwidth]{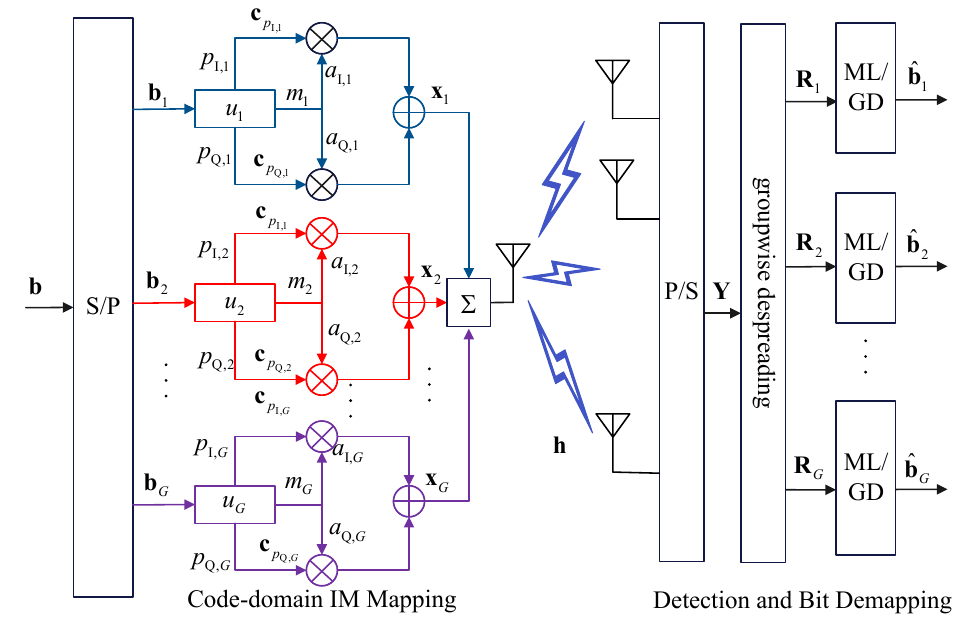}
    \vspace{-0.12in}
    \caption{System block diagram of the proposed CGIM scheme.}
    \label{fig:cgim_system}
    \vspace{-0.12in}
\end{figure}

We consider a single-input multiple-output link with one transmit antenna and
$N_{\mathrm r}$ receive antennas over a frequency-flat channel. The CGIM
transmitter maps each information block onto multiple orthogonal code groups,
as illustrated in Fig.~\ref{fig:cgim_system}.

The transmitter employs a bank of normalized length-$L$ Hadamard orthogonal
spreading codes. From this bank,
$F=GP\leq L$ codes are assigned to $G$ disjoint codebooks, each containing $P$
codes. Let $\mathcal P_g$ denote the global code-index set of group $g$ and
$\mathcal C_g=\{\mathbf c_p:p\in\mathcal P_g\}$ its codebook, where
$\mathbf c_p\in\{-1/\sqrt L,+1/\sqrt L\}^{1\times L}$. The normalized codes satisfy
\begin{equation}
\mathbf c_p\mathbf c_{p'}^{\operatorname T}=\begin{cases}
1,&p=p',\\0,&p\ne p'.
\end{cases}
\label{eq:orthogonality}
\end{equation}

Each CGIM transmission carries the $B$-bit binary stream
$\mathbf b=[\mathbf b_1,\ldots,\mathbf b_G]\in\{0,1\}^{B}$, where
\begin{equation}
B=GB_g=G\left(2\log_2P+\log_2M\right).
\label{eq:spectral_efficiency}
\end{equation}
The stream is divided into $G$ substreams
$\mathbf b_g\in\{0,1\}^{B_g}$, $g=1,\ldots,G$. Within $\mathbf b_g$, the first
and second $\log_2P$-bit segments select
$p_{\mathrm I,g},p_{\mathrm Q,g}\in\mathcal P_g$, respectively, and the
remaining $\log_2M$ bits are mapped to a square $M$-QAM symbol
$s_g=a_{\mathrm I,g}+\mathrm j a_{\mathrm Q,g}$. Here,
$a_{\mathrm I,g},a_{\mathrm Q,g}$ belong to the corresponding $\sqrt M$-PAM
alphabet $\mathcal A$. Thus, each substream is mapped as
$\mathbf b_g\mapsto\langle p_{\mathrm I,g},p_{\mathrm Q,g},s_g\rangle$.
The two code indices are selected independently and may coincide. The group
chip vector is
\begin{equation}
\mathbf x_g=\frac{a_{\mathrm I,g}\mathbf c_{p_{\mathrm I,g}}
+\mathrm j a_{\mathrm Q,g}\mathbf c_{p_{\mathrm Q,g}}}{\sqrt G}.
\label{eq:group_signal}
\end{equation}
Superimposing all groups gives
\begin{equation}
\mathbf x=\frac{1}{\sqrt G}\sum_{g=1}^{G}
\left(a_{\mathrm I,g}\mathbf c_{p_{\mathrm I,g}}
+\mathrm j a_{\mathrm Q,g}\mathbf c_{p_{\mathrm Q,g}}\right).
\label{eq:transmit_signal}
\end{equation}
The factor $1/\sqrt G$ preserves the average block energy
$E_{\mathrm s}=\E[\|\mathbf x\|_2^2]$. Since one
CGIM block conveys $B$ information bits with average energy $E_{\mathrm s}$,
the average bit energy is $E_{\mathrm b}=E_{\mathrm s}/B$.  The resulting spectral efficiency is
$\eta={B}/{L}$ bits per chip. GCIM~\cite{Kaddoum2016GCIM} and QCIM~\cite{Cai2025QCIM} can be viewed as
the $G=1$ and $G=2$ special cases of the proposed CGIM framework,
respectively.

Let $\mathbf h=[h_1,\ldots,h_{N_{\mathrm r}}]^{\operatorname T}$ denote the
channel vector, which remains constant over the $L$ chips of one CGIM
transmission. Its entries are independent and have unit average power. We
consider $h_r\sim\CN(0,1)$ for Rayleigh fading and
$|h_r|^2\sim\operatorname{Gamma}(m,1/m)$ with a uniformly distributed phase
for Nakagami-$m$ fading. The AWGN benchmark is obtained by setting $h_r=1$
for all $r$. The received block is
\begin{equation}
\mathbf Y=\mathbf h\mathbf x+\mathbf N\in\mathbb C^{N_{\mathrm r}\times L},
\label{eq:received_block}
\end{equation}
where the entries of $\mathbf N$ are independent $\CN(0,N_0)$. After channel-phase compensation,
the received sequence at antenna $n$ is
$\mathbf y_n=(h_n^*/|h_n|)(h_n\mathbf x+\mathbf n_n)
=|h_n|\mathbf x+\widetilde{\mathbf n}_n$, where $\mathbf n_n$ is the noise
sequence before compensation and $\widetilde{\mathbf n}_n$ has the same
distribution.

\section{Receiver Design}
\label{sec:receiver}

The orthogonal code-group partition permits the receiver to separate the
superimposed groups before detecting their code indices and PAM levels. For
group $g$, let $\mathbf C_g=[\mathbf c_{p_{g,1}}^{\operatorname T},\ldots,
\mathbf c_{p_{g,P}}^{\operatorname T}]^{\operatorname T}\in\mathbb R^{P\times L}$.
Group-wise despreading yields
\begin{equation}
\mathbf R_g=\mathbf Y\mathbf C_g^{\operatorname T}
=\frac{\mathbf h}{\sqrt G}\left(a_{\mathrm I,g}\mathbf e_{p_{\mathrm I,g}}^{\operatorname T}
+\mathrm j a_{\mathrm Q,g}\mathbf e_{p_{\mathrm Q,g}}^{\operatorname T}\right)
+\widetilde{\mathbf N}_g,
\label{eq:despreading}
\end{equation}
where $\mathbf e_p$ is the $p$th standard basis vector after locally indexing
the $P$ codes in group $g$. The columns of $\widetilde{\mathbf N}_g$ are
independent $\CN(\mathbf0,N_0\mathbf I_{N_{\mathrm r}})$. Hence, orthogonality
removes inter-group interference and confines the I/Q components to their
selected despread columns.

\paragraph{Maximum-Likelihood Detection}
Let $u=(p_{\mathrm I},p_{\mathrm Q},m)$ be a group candidate and
$s_m=a_{\mathrm I,m}+\mathrm j a_{\mathrm Q,m}$. Its noiseless response is
$\overline{\mathbf R}_g(u)=\mathbf h(a_{\mathrm I,m}\mathbf e_{p_{\mathrm I}}^{\operatorname T}
+\mathrm j a_{\mathrm Q,m}\mathbf e_{p_{\mathrm Q}}^{\operatorname T})/\sqrt G$.
The group-wise ML detector is
\begin{equation}
(\widehat p_{\mathrm I,g},\widehat p_{\mathrm Q,g},\widehat m_g)
=\argmin_{u\in\mathcal P_g^2\times\{1,\ldots,M\}}
\|\mathbf R_g-\overline{\mathbf R}_g(u)\|_{\operatorname F}^2.
\label{eq:ml}
\end{equation}

\emph{Proposition 1:} Under mutually orthogonal code groups, independently
applying ML detection to the $G$ despread groups is equivalent to joint ML
detection over the complete CGIM block. The proof is given in
Appendix~\ref{app:group_ml_equivalence}.

With $z_{g,p}=\mathbf h^{\operatorname H}\mathbf r_{g,p}$, removing terms
independent of $u$ reduces the metric to
\begin{equation}
\begin{aligned}
D_g(u)={}&-\frac{2}{\sqrt G}\left[a_{\mathrm I,m}\Re\{z_{g,p_{\mathrm I}}\}
+a_{\mathrm Q,m}\Im\{z_{g,p_{\mathrm Q}}\}\right]
+\frac{\|\mathbf h\|_2^2|s_m|^2}{G}.
\end{aligned}
\label{eq:expanded_metric}
\end{equation}

\paragraph{Greedy Detection}
For square QAM, \eqref{eq:expanded_metric} separates into two PAM subproblems.
Based on this separable structure, we further develop a low-complexity greedy
detection (GD) method. Let
$y_{\mathrm I,g,p}=\Re\{z_{g,p}\}$ and
$y_{\mathrm Q,g,p}=\Im\{z_{g,p}\}$. The GD detector first selects the active
code and then performs conditional PAM slicing:
\begin{subequations}\label{eq:gd_detection}
\begin{align}
\widehat p_{\chi,g}^{\mathrm{GD}}
&=\argmax_{p\in\mathcal P_g}|y_{\chi,g,p}|,
\label{eq:gd_index}\\
\widehat a_{\chi,g}^{\mathrm{GD}}
&=\mathcal Q_{\mathcal A}\!\left(
\frac{\sqrt G\,y_{\chi,g,\widehat p_{\chi,g}^{\mathrm{GD}}}}
{\|\mathbf h\|_2^2}\right),\quad \chi\in\{\mathrm I,\mathrm Q\},
\label{eq:gd_amplitude}
\end{align}
\end{subequations}
where $\mathcal Q_{\mathcal A}(\cdot)$ is the nearest-PAM slicer.

\emph{Corollary 1:} With mutually orthogonal equal-energy codes,
equiprobable square QAM, and white Gaussian noise, GD and group-wise ML give
identical decisions, except for zero-probability ties. The proof is given in
Appendix~\ref{app:gd_ml_equivalence}.

\section{Full-Pair PEP Analysis}
\label{sec:pep}

The identical error statistics of the code groups and I/Q branches reduce the
BER analysis to one real branch. Let $v=(p,i)$ identify code $p$ and normalized PAM amplitude
$\alpha_i=a_i/\sqrt G$, and let $\mathbf b_v$ be its
$b_{\mathrm{br}}=\log_2P+\frac12\log_2M$-bit label. After rotating the
quadrature branch by $-\mathrm j$, either branch can be written as
$\mathbf R_{\mathrm b}=\mathbf h\alpha_i\mathbf e_p^{\operatorname T}
+\mathbf N_{\mathrm b}$. Conditioned on $\mathbf h$, the directed error event
in which the ML detector favors $\widehat v=(\widehat p,j)\ne v$ over the
transmitted state $v$ is
\begin{equation}
\mathcal E_{v\rightarrow\widehat v}=
\left\{\left\|\mathbf R_{\mathrm b}-\mathbf h\alpha_j
\mathbf e_{\widehat p}^{\operatorname T}\right\|_{\operatorname F}^2
\leq
\left\|\mathbf R_{\mathrm b}-\mathbf h\alpha_i
\mathbf e_p^{\operatorname T}\right\|_{\operatorname F}^2\right\}.
\label{eq:pairwise_error_event}
\end{equation}
The squared distance between the two noiseless branch responses is
$\|\mathbf h(\alpha_i\mathbf e_p^{\operatorname T}-\alpha_j
\mathbf e_{\widehat p}^{\operatorname T})\|_{\operatorname F}^2
=\Omega d_{v,\widehat v}^2$, where
\begin{equation}
d_{v,\widehat v}^2=\begin{cases}
(\alpha_i-\alpha_j)^2,&p=\widehat p,\\
\alpha_i^2+\alpha_j^2,&p\ne\widehat p.
\end{cases}
\label{eq:branch_distance}
\end{equation}
and $\Omega=\|\mathbf h\|_2^2$ is the maximal-ratio-combining gain. Substituting
$\mathbf R_{\mathrm b}=\mathbf h\alpha_i\mathbf e_p^{\operatorname T}
+\mathbf N_{\mathrm b}$ into \eqref{eq:pairwise_error_event} gives a real
Gaussian noise projection with variance
$N_0\Omega d_{v,\widehat v}^2/2$. Hence, the conditional PEP (CPEP) is
\begin{equation}
P(v\!\rightarrow\!\widehat v\mid\Omega)
=\Pr\!\left(\mathcal E_{v\rightarrow\widehat v}\mid\mathbf h\right)
=Q\!\left(\sqrt{\frac{\Omega d_{v,\widehat v}^2}{2N_0}}\right).
\label{eq:conditional_pep}
\end{equation}
Craig's representation gives the unconditional PEP (UPEP) as
\begin{equation}
\overline P(v\!\rightarrow\!\widehat v)=\frac{1}{\pi}\int_0^{\pi/2}
\mathcal M_{\Omega}\!\left(-\frac{d_{v,\widehat v}^2}
{4N_0\sin^2\theta}\right)\mathrm d\theta,
\label{eq:average_pep}
\end{equation}
where $\mathcal M_{\Omega}(s)=\E[e^{s\Omega}]$. Under Rayleigh fading,
$|h_r|^2\sim\operatorname{Gamma}(1,1)$ and its MGF is $(1-s)^{-1}$; under
Nakagami-$m$ fading, $|h_r|^2\sim\operatorname{Gamma}(m,1/m)$ and its MGF is
$(1-s/m)^{-m}$~\cite{SimonAlouini2005}. Since
$\Omega=\sum_{r=1}^{N_{\mathrm r}}|h_r|^2$ is a sum of independent branch
gains, $\mathcal M_{\Omega}(s)=\prod_{r=1}^{N_{\mathrm r}}
\mathcal M_{|h_r|^2}(s)$, yielding
\begin{subequations}\label{eq:channel_mgfs}
\begin{align}
\mathcal M_{\Omega}^{\mathrm{Ray}}(s)&=(1-s)^{-N_{\mathrm r}},\\
\mathcal M_{\Omega}^{\mathrm{Nak}}(s)&=(1-s/m)^{-mN_{\mathrm r}},\\
\mathcal M_{\Omega}^{\mathrm{AWGN}}(s)&=e^{N_{\mathrm r}s}.
\end{align}
\end{subequations}
The Nakagami-$m$ result reduces to Rayleigh fading at $m=1$, whereas AWGN has
the deterministic gain $\Omega=N_{\mathrm r}$.

For compact evaluation, the classical approximation
$Q(x)\simeq\frac1{12}e^{-x^2/2}+\frac14e^{-2x^2/3}$ gives
\begin{equation}
\overline P(v\!\rightarrow\!\widehat v)\simeq
\frac1{12}\mathcal M_{\Omega}\!\left(-\frac{d_{v,\widehat v}^2}{4N_0}\right)
+\frac14\mathcal M_{\Omega}\!\left(-\frac{d_{v,\widehat v}^2}{3N_0}\right).
\label{eq:two_exp_pep}
\end{equation}
Finally, with $Q_{\mathrm{br}}=P\sqrt M$ branch states, the bit-weighted
full-pair expression is
\begin{equation}
P_{\mathrm b}^{\mathrm{CGIM}}\lesssim
\frac{1}{Q_{\mathrm{br}}b_{\mathrm{br}}}
\sum_v\sum_{\widehat v\ne v}
d_{\operatorname H}(\mathbf b_v,\mathbf b_{\widehat v})
\overline P(v\!\rightarrow\!\widehat v).
\label{eq:abep}
\end{equation}
Without the approximation in \eqref{eq:two_exp_pep}, replacing $\lesssim$ by
$\leq$ yields the conventional union bound. Equations~\eqref{eq:channel_mgfs}
also show diversity orders $N_{\mathrm r}$ and $mN_{\mathrm r}$ under Rayleigh
and Nakagami-$m$ fading, respectively.

In non-grouped CIM schemes~\cite{Kaddoum2016GCIM,Cai2025QCIM}, joint ML
detection over the complete signal space becomes costly as the transmission
rate increases. In contrast, CGIM enables lossless group-wise ML detection and
reduces the BER analysis to one representative real branch with
$Q_{\mathrm{br}}=P\sqrt{M}$ states.

\section{Spectral Efficiency and Detection Complexity Analysis}
\label{sec:se_complexity}

Table~\ref{tab:code_domain_se} compares the spectral efficiencies of the
considered code-domain IM schemes. Since each transmission occupies $L$ chips,
the spectral efficiency is determined by the number of information bits
conveyed per block. CIM~\cite{Kaddoum2015CIM},
GCIM~\cite{Kaddoum2016GCIM}, and QCIM~\cite{Cai2025QCIM} employ fixed numbers
of code-index branches, whereas CGIM introduces $G$ independently detectable
code groups. For fixed $L$, $P$, and $M$, the spectral efficiency of CGIM
therefore increases linearly with $G$.

\begin{table}[!t]
\caption{Spectral Efficiencies of Code-Domain IM Schemes}
\label{tab:code_domain_se}
\centering
\renewcommand{\arraystretch}{1.12}
\setlength{\tabcolsep}{8pt}
\begin{tabular}{@{}lc@{}}
\toprule
Scheme & Spectral efficiency (bits/chip) \\
\midrule
CIM~\cite{Kaddoum2015CIM}
& $\bigl(\log_2P+\log_2M\bigr)/L$ \\
GCIM~\cite{Kaddoum2016GCIM}
& $\bigl(2\log_2P+\log_2M\bigr)/L$ \\
QCIM~\cite{Cai2025QCIM}
& $\bigl(4\log_2P+2\log_2M\bigr)/L$ \\
Proposed CGIM
& $G\bigl(2\log_2P+\log_2M\bigr)/L$ \\
\bottomrule
\end{tabular}
\end{table}

Detection complexity is measured by the number of real multiplications (RMs)
per CGIM block. Each group contains
$|\mathcal U_g|=P^2M=2^{B/G}$ states. Hence, direct joint ML evaluates
$(P^2M)^G=2^B$ block candidates, whereas the equivalent group-wise
implementation evaluates only $GP^2M=G2^{B/G}$ group candidates. This
reduction is lossless by Proposition~1. Group-wise despreading computes $GP$
length-$L$ real-code
correlations for each of the $N_{\mathrm r}$ receive observations and costs
$2GN_{\mathrm r}PL$ RMs. Reusing $\|\mathbf h\|_2^2$ and
$\mathbf h^{\operatorname H}\mathbf R_g$ across all candidate metrics, the
direct group-wise ML search requires
$G(2N_{\mathrm r}+4N_{\mathrm r}P+3P^2M)$ additional RMs. The separable
detector exploits the independent I/Q PAM subproblems and reduces this count
to $G(2N_{\mathrm r}+4N_{\mathrm r}P+2)$ RMs without changing the ML
decision. Thus, the candidate-dependent complexity decreases from
$\mathcal O(GP^2M)$ to $\mathcal O(GP)$.

\section{Numerical Results}
\label{sec:results}

\begin{figure}[!t]
    \centering
    \subfloat[Rayleigh fading.]{%
        \includegraphics[width=0.95\columnwidth]
        {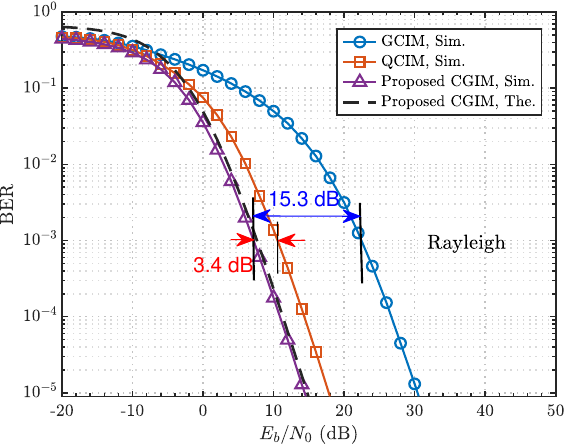}%
        \label{fig:cgim_rayleigh}%
        \vspace{-0.12in}}%
    \vspace{-0.16in}
    \subfloat[Nakagami-$m$ fading ($m=2$).]{%
        \includegraphics[width=0.95\columnwidth]
        {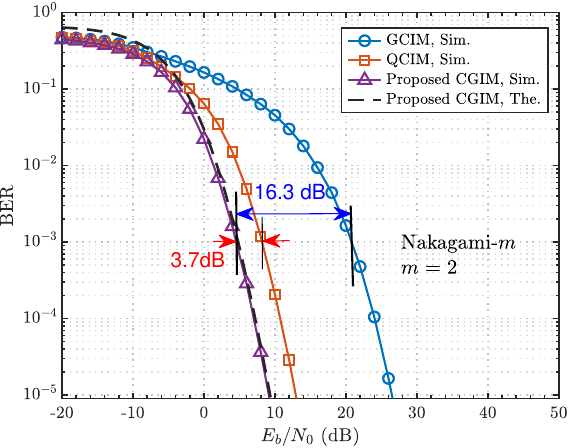}%
        \label{fig:cgim_nakagami}%
        \vspace{-0.12in}}%
    \vspace{-0.16in}
    \subfloat[AWGN.]{%
        \includegraphics[width=0.95\columnwidth]
        {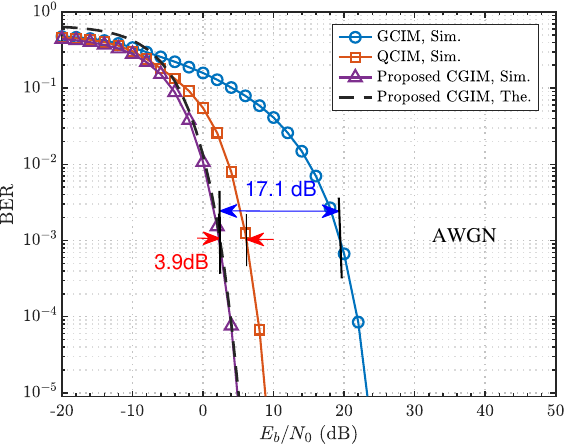}%
        \label{fig:cgim_awgn}%
        \vspace{-0.12in}}
    \vspace{-0.08in}
    \caption{BER comparison of the proposed CGIM with
    GCIM~\cite{Kaddoum2016GCIM} and QCIM~\cite{Cai2025QCIM} over three channel
    models for $L=8$, $N_{\mathrm r}=3$, and $B=16$ bits per transmission.}
    \label{fig:cgim_three_channels}
\end{figure}

The analytical BER results are evaluated against Monte Carlo simulations, and
the proposed CGIM is compared with GCIM~\cite{Kaddoum2016GCIM} and
QCIM~\cite{Cai2025QCIM} at the same transmission rate. The common parameters
are $L=8$, $N_{\mathrm r}=3$, and $B=16$ bits per block. The corresponding
configurations are $(G,P,M)=(1,8,1024)$ for GCIM, $(2,4,16)$ for QCIM, and
$(4,2,4)$ for CGIM. All schemes employ Gray-labeled square QAM, unit average
block energy, and the same $E_{\mathrm b}/N_0$ normalization.

Fig.~\ref{fig:cgim_three_channels} presents the BER results under Rayleigh
fading, Nakagami-$m$ fading with $m=2$, and AWGN in subfigures
\subref{fig:cgim_rayleigh}, \subref{fig:cgim_nakagami}, and
\subref{fig:cgim_awgn}, respectively. Under Rayleigh fading in
Fig.~\ref{fig:cgim_three_channels}\subref{fig:cgim_rayleigh}, CGIM requires
$7.2$ dB at a BER of
$10^{-3}$ and achieves gains of $3.4$ and $15.3$ dB over QCIM and GCIM,
respectively. Since the three schemes use the same number of receive branches,
their simulated curves exhibit the same asymptotic diversity slope, and the
performance differences mainly arise from the distances between the equal-rate
transmit states. For $B=16$ bits per transmission, GCIM and QCIM require
higher-order QAM constellations, whereas CGIM conveys a larger portion of the
information through orthogonal code indices and therefore employs a lower
modulation order. This more balanced allocation between the code-index and
modulation-symbol domains improves the overall distance structure and accounts
for the BER advantage of CGIM.

For Nakagami-$m$ fading in
Fig.~\ref{fig:cgim_three_channels}\subref{fig:cgim_nakagami}, CGIM
requires $4.6$ dB at a BER of $10^{-3}$, corresponding to gains of $3.7$
and $16.3$ dB over QCIM and GCIM, respectively. Under AWGN in
Fig.~\ref{fig:cgim_three_channels}\subref{fig:cgim_awgn}, the required
$E_{\mathrm b}/N_0$
decreases to $2.3$ dB, with gains of $3.9$ and $17.1$ dB over QCIM and
GCIM, respectively. The consistent performance ordering over the three channel
models indicates that the advantage of CGIM mainly stems from its grouped
code-index mapping rather than a specific fading distribution.

The analytical curves closely follow the simulated CGIM results at moderate
and high $E_{\mathrm b}/N_0$. Under Rayleigh and Nakagami-$m$ fading, the
high-SNR slopes agree with the diversity orders $N_{\mathrm r}$ and
$mN_{\mathrm r}$ predicted by \eqref{eq:channel_mgfs}. The low-SNR gap is
mainly due to overlapping pairwise error events in the union bound, and the
analytical results become tighter as the SNR increases.

\begin{figure}[!t]
    \centering
    \includegraphics[width=0.95\columnwidth]
    {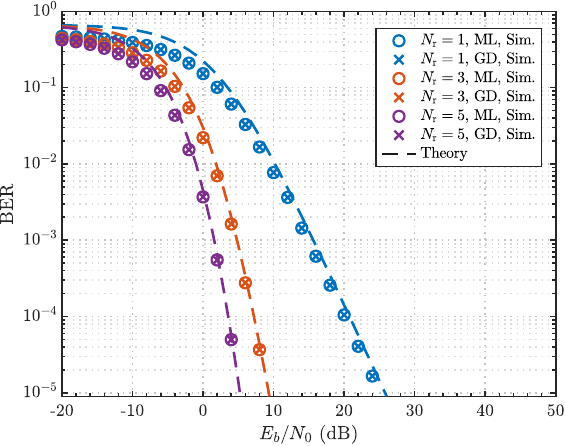}
    \vspace{-0.10in}
    \caption{BER performance of CGIM with ML and GD under Nakagami-$m$
    fading for different numbers of receive antennas, where $m=2$,
    $(G,P,M)=(4,2,4)$, $L=8$, and $B=16$ bits per transmission.}
    \label{fig:cgim_ml_gd_nr}
    \vspace{-0.08in}
\end{figure}

Fig.~\ref{fig:cgim_ml_gd_nr} further compares ML and GD under Nakagami-$m$
fading with $m=2$ for $N_{\mathrm r}\in\{1,3,5\}$. At a BER of $10^{-3}$,
the required $E_{\mathrm b}/N_0$ values decrease from $14.9$ dB for
$N_{\mathrm r}=1$ to $4.6$ and $1.4$ dB for $N_{\mathrm r}=3$ and $5$,
respectively. Increasing $N_{\mathrm r}$ provides larger combining gain and
diversity order, shifting the BER curves toward lower $E_{\mathrm b}/N_0$
with steeper high-SNR slopes. Moreover, the ML and GD curves overlap for all
considered $N_{\mathrm r}$, which agrees with the GD--ML equivalence
established in Appendix~\ref{app:gd_ml_equivalence}.

\section{Conclusion}
\label{sec:conclusion}

This paper proposed CGIM to provide a scalable organization of code-index
resources in spread-spectrum transmission. By partitioning an orthogonal
spreading-code bank into multiple groups, CGIM increases the number of
code-index branches through the group parameter $G$ and enables independent
group-wise despreading and detection. Group-wise ML and low-complexity GD
detectors were developed, where group-wise ML was shown to be equivalent to
joint block-wise ML and GD achieves the same decisions as ML for equal-energy
orthogonal codes and square QAM. A PEP-based BER analysis was further derived
for Rayleigh and Nakagami-$m$ fading, with AWGN considered as a no-fading
benchmark. Numerical results verified the analytical results and showed that,
at the same transmission rate, CGIM achieves better BER performance than GCIM
and QCIM by conveying a larger portion of information through orthogonal code
indices, while the GD detector attains
ML-equivalent BER performance with reduced detection complexity.

\appendices
\section{Proof of Proposition 1}
\label{app:group_ml_equivalence}

\emph{Proof:} Define the orthogonal projector onto the $g$-th code subspace as
$\mathbf P_g=\mathbf C_g^{\operatorname T}\mathbf C_g$. The normalized and
mutually orthogonal codebooks satisfy
$\mathbf P_g\mathbf P_{g'}=\mathbf 0$ for $g\ne g'$. Let
$\mathbf P_\perp=\mathbf I_L-\sum_{g=1}^{G}\mathbf P_g$ project onto the
unused code subspace. Since $\mathbf x_g(u_g)$ lies in the $g$-th subspace,
$\mathbf x_g(u_g)\mathbf P_g=\mathbf x_g(u_g)$ and
$\mathbf x_g(u_g)\mathbf P_{g'}=\mathbf 0$ for $g'\ne g$. Therefore, the
joint ML distance admits the orthogonal decomposition
\begin{equation}
\begin{aligned}
&\left\|\mathbf Y-\mathbf h\sum_{g=1}^{G}\mathbf x_g(u_g)
\right\|_{\operatorname F}^{2}=\left\|\mathbf Y\mathbf P_\perp\right\|_{\operatorname F}^{2}
+\sum_{g=1}^{G}\left\|\mathbf Y\mathbf P_g
-\mathbf h\mathbf x_g(u_g)\right\|_{\operatorname F}^{2}.
\end{aligned}
\label{eq:joint_ml_decomposition}
\end{equation}
The first term is candidate-independent, whereas the $g$-th summand depends
only on $u_g$. Moreover,
$\mathbf C_g\mathbf C_g^{\operatorname T}=\mathbf I_P$, so expressing the
$g$-th projected component by
$\mathbf R_g=\mathbf Y\mathbf C_g^{\operatorname T}$ preserves its Frobenius
distance. It follows that
\begin{subequations}\label{eq:joint_group_ml_equivalence}
\begin{align}
\widehat u_g^{\mathrm{GML}}
&=\argmin_{u_g\in\mathcal U_g}
\left\|\mathbf R_g-\overline{\mathbf R}_g(u_g)
\right\|_{\operatorname F}^{2},
\label{eq:individual_group_ml}\\
\widehat{\mathbf u}_{\mathrm{JML}}
&=\left(\widehat u_1^{\mathrm{GML}},\ldots,
\widehat u_G^{\mathrm{GML}}\right).
\label{eq:joint_group_ml_solution}
\end{align}
\end{subequations}
Hence, group-wise ML and joint block-wise ML produce identical decisions.
\hfill$\blacksquare$

\section{Proof of Corollary 1}
\label{app:gd_ml_equivalence}

\emph{Proof:} Put $H=\|\mathbf h\|_2^2$. For either real branch with correlator output $y$,
the candidate-dependent ML metric satisfies
\begin{equation}
D(p,a)=\frac{H}{G}a^2-\frac{2a}{\sqrt G}y
=\frac{H}{G}\left(a-\frac{\sqrt G}{H}y\right)^2-\frac{y^2}{H}.
\label{eq:appendix_gd_branch_metric}
\end{equation}
Thus, for a fixed code index, its minimizer is precisely the nearest-PAM
slicer in \eqref{eq:gd_amplitude}. After minimizing over $a$, define the
branch score and consider $t_2>t_1\geq0$:
\begin{subequations}\label{eq:appendix_gd_monotonicity}
\begin{align}
\Lambda(t)&=\max_{a\in\mathcal A}
\left\{\frac{2at}{\sqrt G}-\frac{H}{G}a^2\right\},
\label{eq:appendix_gd_score}\\
\Lambda(t_2)&\geq\Lambda(t_1)
+\frac{2a_1}{\sqrt G}(t_2-t_1)>\Lambda(t_1),
\label{eq:appendix_gd_strict_increase}
\end{align}
\end{subequations}
where $a_1>0$ is a maximizer of $\Lambda(t_1)$. Symmetry of the PAM alphabet
gives $\Lambda(-t)=\Lambda(t)$, while
\eqref{eq:appendix_gd_strict_increase} shows that the score is strictly
increasing in $|t|$. Hence, ML selects the largest $|y|$, as in
\eqref{eq:gd_index}. Additivity of the I/Q metrics proves the
GD--group-wise-ML equivalence, and Proposition~1 extends it to joint block-wise
ML. \hfill$\blacksquare$

\bibliographystyle{IEEEtran}
\bibliography{Reference}

\end{document}